\documentclass{article}
\usepackage{graphicx}
\usepackage{amsmath,amssymb}
\usepackage{graphics,color,array,calc,rotating,epsfig,psfrag}
\usepackage{hyperref}
\numberwithin{equation}{section}
\usepackage{cite}
\usepackage{bm}
\usepackage{dcolumn}
\usepackage{float}
\usepackage{subfigure}
\usepackage{amsfonts}
\usepackage[mathscr]{eucal}
\usepackage{helvet}
\newcommand\T{\rule{0pt}{2.6ex}}       
\newcommand\B{\rule[-1.2ex]{0pt}{0pt}} 
\def\be{\begin{equation}} \def\ee{\end{equation}}
\def\bea{\begin{eqnarray}} \def\eea{\end{eqnarray}}

\newcommand\prt{\partial}

\newcommand{\nn}{\nonumber}
\newcommand{\RN}[1]{%
  \textup{\uppercase\expandafter{\romannumeral#1}}%
}
\begin{document}
\baselineskip 18pt%
\begin{titlepage}
\vspace*{1mm}%
\hfill%
\vspace*{15mm}%
\hfill
\vbox{
    \halign{#\hfil         \cr
        IPM/P-2021/51  \cr
          } 
      }  
\vspace*{20mm}
\begin{center}
{\Large {\bf Complexity growth in Gubser-Rocha models with momentum relaxation}}
 \end{center}
\vspace*{5mm}
\begin{center}
{H. Babaei-Aghbolagh$^{a}$, Davood Mahdavian Yekta$^{b}$,  Komeil Babaei Velni$^{c,d}$, H. Mohammadzadeh$^{a}$}\\
\vspace*{0.2cm}
{\it
$^{a}$Department of Physics, University of Mohaghegh Ardabili,
P.O. Box 179, Ardabil, Iran\\
$^{b}$Department of Physics, Hakim Sabzevari University, P.O. Box 397, Sabzevar, Iran\\
$^{c}$Department of Physics, University of Guilan, P.O. Box 41335-1914, Rasht, Iran\\
$^{d}$School of Physics,
Institute for Research in Fundamental Sciences (IPM), P.O. Box 19395-5531, Tehran, Iran\\
}

 \vspace*{0.5cm}
{E-mails: {\tt h.babaei@uma.ac.ir, d.mahdavian@hsu.ac.ir, babaeivelni@guilan.ac.ir,  mohammadzadeh@uma.ac.ir}}
\vspace{1cm}
\end{center}

\begin{abstract}
The Einstein-Maxwell-Axion-Dilaton (EMAD) theories, based on the Gubser-Rocha (GR) model, are very interesting in holographic calculations of strongly correlated systems in the condensed matter physics. Due to the presence of spatially dependent massless axionic scalar fields, the momentum is relaxed and we have no translational invariance at finite charge density. It would be of interest to study some aspects of quantum information theory for such systems in the context of $AdS/CFT$ where EMAD theory is a holographic dual theory. For instance, in this paper we investigate the complexity and its time dependence for charged $AdS$ black holes of EMAD theories in diverse dimensions via the complexity equals action (CA) conjecture. We will show that the growth rate of the holographic complexity violates the Lloyd's bound at finite times. However, as shown at late times, it depends on the strength of momentum relaxation and saturates the bound for these black holes.

\end{abstract}

\end{titlepage}

\section{Introduction}

Strongly correlated systems in the condensed matter physics include some interesting phenomena which are not easy to analyze theoretically, however in the context of holography \cite{tHooft:1993dmi,Susskind:1994vu}, there are some practical ways which map them to the dual weakly interacting systems  \cite{Hartnoll:2009sz,Hartnoll:2016apf,Zaanen,Herzog:2009xv,Sachdev:2011wg}. For instance, linear T-resistivity in the strange metals has a remarkable degree of universality which does not observed in general Fermi liquids with quadratic temperature dependence for resistivity \cite{Hartnoll:2009ns,Charmousis:2010zz,Davison:2013txa,Lucas:2014zea}, or the divergent conductivity of systems with momentum relaxation due to a broken translational symmetry which occurs in more realistic condensed matter materials \cite{Liu:2012tr,Ling:2013aya,Ling:2014bda,Mozaffara:2016iwm,Cremonini:2018xgj,Cremonini:2019fzz}.
In this way, the GR models \cite{Gubser:2009qt} and the EMAD theories \cite{ Gouteraux:2014hca,Caldarelli:2016nni}, based on the GR model, are of the most interesting theories which provide nice areas for the study of electric and magnetic transport phenomena in a strongly-coupled system. The EMAD models can also be dilatonic generalization of EMA theories proposed in \cite{Andrade:2013gsa}.

Apart from this holographic approach of solving strongly coupled field theories in condensed matter physics, the $AdS/CFT$ correspondence \cite{Maldacena:1997re,Witten:1998qj,Gubser:1998bc,Itzhaki:1998dd,Aharony:1999ti} has been far more deep and revealing than merely providing a classical geometrical computational tool for strongly coupled field theory phenomena. Thinking about how field theory codes various phenomena on the gravity side has led to the recognition of various concepts from the quantum information. Information geometry \cite{Dennis:2001nw}, Von-Neumann \cite{Ryu:2006bv} and Renyi entropy \cite{Dong:2016fnf}, mutual information \cite{Headrick:2007km}, tensor networks \cite{Swingle:2009bg,Czech:2015kbp}, computational complexity \cite{Susskind:2014rva}, fidelity or relative entropy \cite{Lin:2014hva} and quantum error correcting codes \cite{Almheiri:2014lwa} are only to name a few.
Among them, entanglement entropy has been the most fundamental thing to study as it measures the correlation between two subsystems \cite{Ryu:2006ef,Hubeny:2007xt,Nishioka:2009un}. However, entanglement entropy may not be enough to probe the degrees of freedom interior the black holes, since the volume of black holes continues growing even if spacetimes reach thermal equilibrium\cite{Susskind:2014moa}. This motivates the introduction of holographic complexity in the quantum information theory \cite{Susskind:2014rva,Stanford:2014jda}. As a measure, quantum complexity describes how many simple elementary gates (unitary operators) are needed to obtain a particular state from some chosen reference state \cite{watrous2009quantum,Aaronson:2016vto,Arona}. (For more definitions in quantum field theories see e.g. \cite{Jefferson:2017sdb,Chapman:2017rqy,Hashimoto:2018bmb,Chapman:2018hou})

In the context of $AdS/CFT$ correspondence, two holographic prescriptions have been suggested to compute the complexity; the ``complexity=volume'' (CV) conjecture  \cite{Susskind:2014rva,Stanford:2014jda} and the ``complexity= action'' (CA) conjecture \cite{Brown:2015bva,Brown:2015lvg}. In the CV conjecture, the complexity of a state is proportional to the volume of Einstein-Rosen bridge which connects two boundaries of an eternal black hole, i.e. $\mathcal{C}_{V}= \mathcal{V}/G\,\ell$ where $G$ is the Newton's constant and $\ell$ is an arbitrary length scale, while in the CA picture, it is proportional to the bulk action evaluated in a certain spacetime region known as the Wheeler-De Witt (WDW) patch, i.e. $\mathcal{C}_A =I_{WDW}/\pi\hbar$.
By construction, the CA proposal is devoid of the ambiguity associated with arbitrary length scale $\ell$ appeared in the CV picture. There are two conceptual features here; $i)$ A gravity system for which the holographic complexity has been studied extensively is the eternal two-sided $AdS$ black hole \cite{Maldacena:2001kr} which is dual to a thermofield double state in the dual boundary field theory,
\be\label{TFD} |\psi_{TFD}\rangle=\frac{1}{\sqrt{Z}}\sum_j e^{-{E_j}/({2T})} e^{-i E_{j}(t_{L}+t_{R})}|E_j\rangle_L |E_j\rangle_R\,,\ee
where $L$ and $R$ refer to the left and right entangled regions of the two sided black hole. The entanglement is due to the Einstein-Rosen bridge that connects the left and right boundary $CFT$s.
$ii)$ For $AdS$ black holes of mass $M$ in the CA conjecture, the late time behavior of the change rate of complexity reaches a constant value
\be\label{LB} \dot{\mathcal{C}}_A\leq {2M},\ee
where in the rest of the paper we assume that $\pi\hbar=1$. This is often referred to the Lloyd's bound \cite{Lloyd}, which in the case of charged and rotating black holes is generalized respectively to \cite{Brown:2015lvg,Cai:2016xho}
 \be \label{cjlt}
 \dot{\mathcal{C}}_A\leq2\big[(M-\mu Q)-(M-\mu Q)_{gs}\big],\qquad
 \dot{\mathcal{C}}_A\leq2\big[(M-\Omega J )-(M-\Omega J )_{gs}\big].
 \ee
Here,``$gs$'' denotes the ground state of the black hole.
The hope that the holographic complexity might provide useful information about the spacetime structure behind the horizon has led to intense investigation of the CA and CV proposals in various gravity theories, probing their structure and properties \cite{Brown:2017jil}-\cite{Camargo:2019isp}.

In the light of the above discussion, it would be of interest to study the complexity for charged $AdS$ black holes in EMAD theories which may be examined in holographic models of quantum information theory and quantum computation. Another motivation is  to investigate that whether the contribution of different fields in EMAD theory can affect the Lloyd's bound for the corresponding black holes here, or whether in general there is a violation for this bound or not. In particular, we will compute the holographic complexity and its time evolution for charged single-horizon solutions of 4, 5, and $(d+1)$-dimensional EMAD theories using the CA conjecture. We study both the early and late time limit for the growth rate of complexity. This may not only allow us to further constrain the validity of these holographic conjectures but also provide other examples in the growing list of literature where the Lloyd's bound can be explicitly violated.

It has been shown in \cite{Brown:2015bva} that for small Reissner-Nordstrom $AdS$ black holes \footnote{In theoretical physics, an extremal black hole is a black hole with the minimal possible mass that can be compatible with a given charge and angular momentum. In other words, this is the smallest possible black hole that can exist while rotating at a given fixed constant speed with some fixed charge.} of mass $M$ and charge $q$ in EM theory, the Lloyd's bound at late time is given by $\dot{\cal{C}}_A=2 M - \mu q$, while for charged $AdS$ black holes in a simple class of EMD theories one has $\dot{\cal{C}}_A=2 M -\mu q - C$ where $C$ is a constant term \cite{Cai:2017sjv}. On the other hand, we have shown in \cite{Yekta:2020wup} that for neutral black branes in EMA theory, the momentum relaxation term does not explicitly change the late time limit $\dot{\mathcal{C}}_A=2M$. Acually, this fact is another main motivation for us to investigate in this paper whether the momentum relaxation term would correct the Lloyd's bound in EMAD theories or not.

The outline of the paper is as follows: In section \ref{sec2}, we review the holographic complexity in the CA picture. We evaluate the action on a WDW patch that has null boundary surfaces in addition to time/space-like boundaries. In this sense, we should consider some extra actions to remove the ambiguities for null surfaces \cite{Lehner:2016vdi}. In section \ref{sec3}, we calculate the complexity growth of charged $AdS$ black holes for EMAD holographic models in diverse dimensions. We also provide a numerical study on $\dot{\mathcal{C}}_A$ to find how the Lloyd's bound is violated. In section \ref{sec4}, we summarize our
results and present a discussion about computed bounds in different theories.
\section{Holographic Complexity via CA conjecture}\label{sec2}

Our main interest in this paper is to investigate the time dependency of holographic complexity, however, there are two distinct counting methods of the complexity growth rate in the CA picture. The first proposal by Brown et al \cite{Brown:2015bva,Brown:2015lvg} puts forward to calculating $\dot{\mathcal{C}}_A$ at late times while in the latter proposal by Lehner et al \cite{Lehner:2016vdi}, the complexity is evaluated as well as its time evolution. In general, they are not equivalent but give the identical very late time results \cite{Jiang:2019pgc}. The main difference between two methods is that in the former, there is no way to violate the Lloyd's bound, while as we have shown the bound is violated from above in the latter. Therefore, we use the second approach to study the holographic complexity and its time evolution for charged $AdS$ solutions of EMAD theories. The essential ingredient in the CA conjecture is to evaluate the action on a WDW patch \cite{Brown:2015bva,Brown:2015lvg}. However, we follow the method in Ref. \cite{Carmi:2017jqz} where not only the action on the WDW patch includes the bulk and the Gibbons-Hawking-York (GHY) boundary terms \cite{York:1972sj,Gibbons:1976ue}, but also embraces boundary segments of joint terms due to the intersection of time-like, space-like, and null boundaries \cite{Hayward:1993my}. This will be the general strategy that will be followed in the next section.

The bulk actions for holographic EMAD models are denoted by $I_{bulk}$ and essentially are different in diverse dimensions. However, the contribution of GHY surface terms are the same for all of them as follows
\be\label{surf} I_{surf}=\frac{1}{8\pi G}\int_{\mathcal{B}} d^{d} x \sqrt{h} K-\frac{1}{8\pi G}\int_{\mathcal{B'}} d\lambda\, d^{d-1}\theta \sqrt{\gamma}\,\kappa,\ee
where $K$ is the extrinsic curvature of the time-like or space-like boundaries ($\mathcal{B}$) and $\kappa$ is extrinsic curvature of the null segments ($\mathcal{B}'$). Here, $h_{\mu\nu}$ and $\gamma_{\mu\nu}$ are the induced metrics on $\mathcal{B}$ and $\mathcal{B}'$, respectively, so that $K_{\mu\nu}=-h_{\mu}^{\rho} h_{\nu}^{\sigma} \nabla_{(\rho}n_{\sigma)}$ and $\kappa=-N_{\mu} k^{\nu} \nabla_{\nu} k^{\mu}$ where $n_{\mu}$ and $k_{\mu}$ are outward pointing unit normal vectors on them. $\lambda$ parameterizes the null generator of the null boundary such that for affinely parametrization $\kappa=0$ \cite{Lehner:2016vdi}.

\begin{figure}[H]
\centering
\includegraphics[width=9cm,height=6cm]{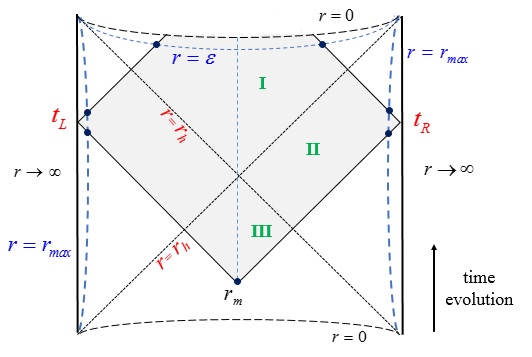}
\caption{Penrose diagram of the WDW patch for $AdS$ black holes. $r_0$ is the physical singularity and $r={\varepsilon}$ and $r=r_{max}$ are the IR and UV cut off surfaces, respectively.}
\label{f1}
\end{figure}

The joint actions are given by
\be\label{joint} I_{jnt}=\frac{1}{8\pi G}\int_{\Sigma}  d^{d-1}x \sqrt{\sigma}\,\eta+ \frac{1}{8\pi G}\int_{\Sigma '} d^{d-1}x \sqrt{\sigma}\,a,\ee
where $\eta$ corresponds to the intersection of time-like or space-like boundaries while $a$ is related to intersecting of null surfaces. As shown in Fig.~(\ref{f1}), the WDW patch have intersections of null surfaces with time-like or space-like boundaries at $r={\epsilon}$ and $r=r_{max}$ but since they have no time dependence \cite{Chapman:2016hwi}, it is not necessary to consider them here. Thus, only the contribution of null meeting point $r_{m}$ remains for joint action.

There is also a counterterm action for the null surfaces as
\be\label{cterm} I_{ct}=\frac{1}{8\pi G} \int_{\mathcal{B'}} d\lambda \,d^{d-1}\theta \sqrt{\gamma}\, \Theta \log{(\ell_{c}\Theta)}\,,\ee
 which is introduced to ensure reparametrization invariance on the null boundaries and does not change the variational principle. Indeed, the parameter expansion $\Theta$ is the relative rate of change of the cross-sectional area of a bundle of null generators  which is denoted by
\be\label{amb}\Theta=\prt_{\lambda} \log{\sqrt{\gamma}},\ee
and $\ell_{c}$ is an arbitrary length scale (for more discussions see e.g.\cite{Lehner:2016vdi}). Therefore, we should investigate the following total action on the WDW patch
\be\label{tact} {I}_{WDW}=I_{bulk}+I_{surf}+I_{jnt}+I_{ct}\,.\ee

In Fig.~(\ref{f1}) we have plotted the Penrose diagram of a two-sided $AdS$ black hole with single horizon $r_h$. The grey shaded region is the WDW patch bounded by the light sheets sent from the two asymptotic symmetric time slices $t_L$ and $t_R$ such that $t_L=t_R\equiv t/2$. The WDW patch includes two equivalent sectors where the future and past null boundaries of the right sector are given respectively by
\be\label{BB} {t_L}=r_{}^{*} (r)-{r}_{\infty}^{*}\,,\qquad -{t_R}=r_{}^{*} (r)-{r}_{\infty}^{*}\,.\ee
In general, $r^{*}$ is a tortoise coordinate defined as $dr^{*} (r)={d r}/{f ( r)}$ upon which the new coordinates $v=t+r^{*} (r)$ and $u=t-r^{*} (r)$ are defined and $\lim_{r \rightarrow \infty} r_{}^{*} (r) = {r}_{\infty}^{*}$. Also $r_m$ is the meeting point of null boundaries before reaching the past singularity at critical time $t_c=2({r}_{\infty}^{*}-r_{}^{*} (0))$.

\section{Holographic Complexity for EMAD theories} \label{sec3}
In this section we consider the holographic complexity and its time evolution for three families of holographic EMD theory based on the GR model in the presence of momentum relaxation term constructed from axionic scalar fields. The general bulk action of EMAD theories in arbitrary dimensions is given by \cite{Gouteraux:2014hca}
\be\label{pact}
I = {1 \over 16\pi G_{d+1}} {\int_{}^{}} d^{d+1}x   \sqrt{-g}  \left[  R -  \frac{1}{2}(\partial_\mu\phi)^2 + V(\phi) -  \frac{1}{4} Z(\phi) F_{\mu\nu}^2  -\frac{1}{2} \sum_{I=1}^{d-1}(\partial_\mu \psi_{I})^2  \right],
\ee
where $G_{d+1}$, $\phi(r)$, and $F_{\mu\nu}\!=\!\prt_{\mu}A_{\nu}-\prt_{\nu}A_{\mu}$ are the ${d+1}$-dimensional Newton's constant, dilaton field and field strength of $U(1)$ gauge field $A_{\mu}$, respectively. Also, $V(\phi)$ is a dilatonic scalar potential and $Z(\phi)$ is a coupling function between the Maxwell and dilaton fields. The axionic scalar fields enter the bulk action only through the kinetic term $\prt_{\mu}\psi_{I} $ and the sources are linear in the boundary, i.e. $\psi^{(0)}_{I} \propto \beta_{I i} x^i$ \cite{Andrade:2013gsa}, where $\beta$ represents the strength of the momentum relaxation for which
\be \beta^2\equiv\frac{1}{2}\sum_{i=1}^{d-1} \vec{\beta}_{i}\cdot \vec{\beta}_{i}, \qquad \vec{\beta}_{i}\cdot \vec{\beta}_{j}=\sum_{I=1}^{d-1}{\beta}_{Ii}\, {\beta}_{I j} =\beta^2 \delta_{i j} \qquad \forall \, i, j=1,2,...,d-1.\ee
The field equations of the action (\ref{pact}) are given by
\bea
\label{eqs}
&R_{\mu\nu}-\frac{V(\phi)}{d-1}g_{\mu\nu}-\frac12 \prt_{\mu} \phi \,\prt_{\nu} \phi -\frac12 Z(\phi) {F_{\mu}}^{\rho}F_{\nu\rho}-\frac{1}{4(d-1)} g_{\mu\nu} Z(\phi) F^2-\frac12  \sum_{I=1}^{d-1} {\prt_{\mu} \psi_I \,\prt_{\nu} \psi_I}=0,\nn\\
&\nabla_{\mu}\left(Z(\phi) F^{\mu\nu}\right)=0,&\\
&\Box \psi_I=0,\qquad I=1,..,d-1,&\nn\\
&\Box\phi+\frac{d V(\phi)}{d\phi}-\frac14 \frac{d Z(\phi)}{d\phi} F^2=0.&\nn
\eea
In general, the analytical charged solutions of these equations are described by the following ansatz
\be\label{metric}
ds^2 = e^{2A(r)} \left(-h(r) \, dt^2 + d\vec{x}^2\right) +  \frac{e^{2B(r)}}{h(r)} dr^2,
\ee
where $ d\vec{x}^2$ is the line element of $(d-1)$-dimensional spatial flat space \footnote{To describe the line element of the black holes, in this paper, we use the convention in Ref.~\cite{Kim:2016dik} in which the holographic renormaliation of EMD theories have been considered.}. In this regard, we evaluate the total action (\ref{tact}) on the WDW patch shown in Fig.~(\ref{f1}) for each theory. We also determine the Lloyd's bound on the rate of complexity at late times.

Since we are going to analyze the holographic complexity of charged black holes in asymptotically $AdS$ space, it is expected that one should consider the causal structure of Reissner-Nordstrom $AdS$ black hole given in figure 14 of \cite{Chapman:2016hwi} in which the WDW patch have two intersecting points correspond to either future or past null boundaries where at the late time coincide with the inner and outer horizons respectively. But, as will be shown for charged black holes in this paper, the inner Cauchy horizon is replaced by a curvature singularity at $r=0$, therefore, the causal structure is described by the one for $AdS$-Schwarzschild black hole as depicted in Fig.~(\ref{f1}).

\subsection{$AdS_4$ black holes}\label{sec31}

In four dimensions the EMAD theory is constructed from two parts: an EMD theory obtained from dimensional reduction of $AdS_4\times S^{7}$ in $M$- theory to triple intersection of $M5$-branes that has the three-equal-charge black hole \cite{Gubser:2000mm} and a kinetic term from two massless axionic scalar fields $\psi_{I} $ with $I=1,2$ \cite{Gouteraux:2014hca}. This theory is described by the following bulk action
\be\label{act4}
  {I}_{bulk} \!= \!{1 \over 16\pi G_4}\int d^4x \sqrt{-g}\left[ R - {1 \over 4} e^{\phi} F_{\mu\nu}^2 -
   {3 \over 2} (\partial_\mu\phi)^2 + {6 \over L^2} \cosh \phi -\tfrac{1}{2} \sum_{I=1}^2 {(\partial_\mu\psi_I)}^2 \right] \!= \!{1 \over 16\pi G_4}\int d^4x \sqrt{-g} \mathcal{L}_1, \,
\ee
where $\mathcal{L}_1$ is the Lagragian density introduced for later convenience. For $\phi(r)=0$, the model reduces to the action in Ref.~\cite{Andrade:2013gsa,Davison:2015bea} for which we have extensively considered its holographic complexity in Ref.~\cite{Yekta:2020wup}. Holographic calculations of this model have been studied in Refs.~\cite{Zhou:2015qui,Kim:2017dgz,Jeong:2018tua,Jeong:2019zab}. Substituting the ansatz (\ref{metric}) in the equations of motion, we obtain
\bea \label{p1}
&&A(r)=-B(r)=\log\frac{r}{L}+\frac34\log\left(1 + \frac{Q}{r}\right),\quad  \phi (r) =\frac{1}{2}\log\left(1 + \frac{Q}{r}\right),\nn\\
&&h(r)= 1-  \frac{L^4  \beta^2}{2 (Q + r)^2} -  \frac{(Q + r_h)^3 }{(Q + r)^3}\bigl(1 -  \frac{L^4 \beta^2}{2 (Q + r_h)^2}\bigr),\quad \psi_I =\beta_{I i} x^i,\\
&&  A_\mu dx^\mu =A_{t}(r) dt,\quad A_{t}(r)=\frac{r- r_h}{Q + r}\sqrt{\frac{3Q (Q + r_h)}{L^2} \bigl(1 -  \frac{L^4  \beta^2}{2 (Q + r_h)^2}\bigr)}, \nn
\eea
where $Q$ represents a charge parameter and $L$ is a length scale that henceforth we set $L=1$. We can recast the solution (\ref{metric}) in the form of a general $AdS$ black hole as
\be
 \label{ads40} ds^2 =-f(r)\, dt^2 +{U(r)} d\vec{x}^2 + \frac{1}{f(r)} dr^2,
\ee
in which
\be \label{fp}
f(r)= \frac{r^{1/2} (r- r_h ) \left(3 Q^2 + r_h^2 + r_h r + r^2 + 3 Q (r_h + r)\right)}{(Q + r)^{3/2}}-\frac{ r^{1/2} (r-r_h )\beta^2 }{2 (Q + r)^{3/2}}, \quad U(r)=r^{1/2} (Q + r)^{3/2}.
\ee
This expression of $f(r)$ explicitly shows that the solution has two singularities at $r=r_h$ and $r=0$ where the first corresponds to the event horizon and the latter is the spacetime singularity. Substituting the functions (\ref{fp}) and parameters (\ref{p1}) in to the Lagrangian (\ref{act4}), we have
\be
\sqrt{-g} {\mathcal{L}}_1= -  \frac{3 \bigl(Q^4 + 4 r^4 + Q^3 (-3 r_h + 10 r) - 3 Q^2 (r_h^2 - 6 r^2) -  Q (r_h^3 - 14 r^3)\bigr)}{2 (Q + r)^2}- \frac{3 Q (Q + r_h) \beta^2}{4 (Q + r)^2} .
\ee

As alluded in the introduction, it is costum to write the late time behavior of $\dot{\mathcal{C}}_A$ in terms of the physical parameters of black holes, thus we should calculate them for each solution. The mass of the black hole is given by
\be\label{mass4}
 M={ V_2 \over 16\pi G_4} 2\omega,\qquad \omega= (Q + r_h)^3\left(1-\frac{\beta^2}{2(Q + r_h)^{2}}\right),
\ee
where $\omega$ is a mass parameter and $V_2$ is the dimensionless volume of 2-dimensional spatial geometry $\vec{x}$. In the zero dissipation $\beta\rightarrow 0$, the mass parameter reduces to $\omega=Q^3$ for extremal limit $r_h=0$ in Ref.~\cite{Gubser:2009qt}. Also, in the limit $Q\rightarrow 0$, one achieves the mass of neutral black hole given by Eq.~(2.36) in Ref.~\cite{Yekta:2020wup}. The chemical potential and black hole charge are obtained from $A_t$ in (\ref{p1}) as
\be\label{cpcd}
\mu = \sqrt{3Q (Q + r_h) \left(1 -  \frac{ \beta^2}{2 (Q + r_h)^2}\right)},\qquad
q =\frac{V_2}{16\pi G_4}\sqrt{3Q (Q + r_h)^3 \left(1 -  \frac{ \beta^2}{2 (Q + r_h)^2}\right)},
\ee
where the chemical potential can be read off from the asymptotic value of the electric potential $A_{t}(r\rightarrow \infty)$ and $q$ is computed from the Gauss's law.
The temperature and entropy of the black hole are obtained from
\be\label{ST}
T=\frac{f(r)'}{4\pi}\Big|_{r=r_h}=\frac{r_h^{1/2} \bigl(6 (Q + r_h)^2 -  \beta^2\bigr)}{8 \pi (Q + r_h)^{3/2}},\quad S=\frac{A_h}{4G_4}\Big|_{r=r_h}=\frac{V_2}{4 G_4} r_h^{1/2} (Q + r_h)^{3/2},
\ee
where prime is the derivative with respect to $r$ and $A_h=V_2 U(r_h)$ is the area of the event horizon, such that
\be\label{STp}
T S=\frac{V_2}{8 \pi G_4} \frac{r_h}{4}  \left(6 (Q + r_h)^2 -  \beta^2\right).
\ee
\subsubsection{The growth rate of complexity}
We will now proceed to compute the time evolution of the holographic complexity via the CA proposal. Due to the left-right symmetry of the WDW patch for planar $AdS$ black hole in Fig.~(\ref{f1}) we only work on the right side for times $t>t_c$ and then multiplied the final result by a factor of 2 for symmetric left side. The contribution of the bulk action (\ref{act4}) is given by
\bea \label{bulk1} I_{bulk}&\!\!\!\!\!=\!\!\!\!\!&2\,\, \left(I_{bulk}^{\RN{1}}+I_{bulk}^{\RN{2}}+I_{bulk}^{\RN{3}}\right)\nn\\
&\!\!\!\!\!=\!\!\!\!\!&\frac{V_2}{8\pi G_4}\left[\int_{\epsilon}^{r_h}\left(\frac{t}{2}+{r}_{\infty}^{*}-r_{}^{*} (r)\right)+2\int_{r_h}^{r_{max}}\left({r}_{\infty}^{*}-r_{}^{*} (r)\right)+\int_{r_m}^{r_h}\left(-\frac{t}{2}+{r}_{\infty}^{*}-r_{}^{*} (r)\right) \right]\, \sqrt{-g}\, \mathcal{L}_1\, dr,\nn\\
&\!\!\!\!\!=\!\!\!\!\!&I_{bulk}^{0}+\frac{V_2}{8\pi G_4} \int_{\epsilon}^{r_m}\left(\frac{t}{2}+{r}_{\infty}^{*}-r_{}^{*} (r)\right) \, \sqrt{-g}\, \mathcal{L}_1\, dr,
\eea
where $I_{bulk}^{0}$ is the time independent part, therefore we have
\be \label{br1}
\frac{d I_{bulk}}{dt }= \frac{V_2}{16\pi G_4 }\left[-3 Q^2 r -  \frac{9}{2} Q r^2 - 2 r^3 -  \frac{3Q (Q+r_h)^3}{2 (Q + r)} + \frac{3 Q (Q+r_h)\beta^2}{4 (Q + r)}\right]_{\epsilon}^{r_{m}}.
\ee

According to Fig.~(\ref{f1}), the boundary action includes two distinguished parts: a time-like surface at the IR cut off and a space-like surface at the UV cut off point. The extrinsic curvature for the ansatz (\ref{ads40}) is as fallows
 \be \label{EX} K=\frac{1}{2}\left(\frac{\prt_r f(r)}{\sqrt{f(r)}}+2 \sqrt{f(r)}\, \prt_r{\ln\left(U(r)\right)}\right), \ee
 where the normal vectors corresponding to these surfaces are given by
\be \label{NV} r={\epsilon}:\quad {\bf t}=t_{\mu} dx^{\mu}=-\frac{dr}{\sqrt{-f(\epsilon)}},\qquad \qquad r=r_{max}:\quad {\bf s}=s_{\mu} dx^{\mu}=\frac{dr}{\sqrt{f(r_{max})}}.\ee
 Again by supposing a factor of 2 for the left sector of patch we have
 \be\label{surf3} I_{surf}=2\,(I_{surf}^{r=\epsilon}+I_{surf}^{r=r_{max}})=I_{surf}^0+\frac{V_2}{4\pi G_4}\sqrt{h} K\,\left(\frac{t}{2}+{r}_{\infty}^{*}-r_{}^{*} (r)\right)\Big|_{r=\epsilon}.
 \ee
Here $I_{surf}^0$ includes the contribution of the UV cutoff surfaces whish is independent of time, thus the time evolution of the GHY surface term is
\bea
 \frac{{d I_{surf}}}{dt}&\!\!\!=\!\!\!&\frac{V_2}{8\pi G_4}\Big[-\frac{3 \bigl(Q^3 (3 r_h - 5 r) + 3 Q^2 (r_h^2 + 2 r_h r - 5 r^2) + Q (r_h^3 + 6 r_h^2 r - 13 r^3) + 2 r (r_h^3 - 2 r^3)\bigr)}{4 (Q + r)} \nn\\
 &&\quad\qquad-\frac{  (8 r^2+5 Q r- 6 r_h r-3 Q r_h  )\beta^2}{8 (Q + r)}\Big]_{r=\epsilon}.
 \eea
 
It has been shown in Ref. \cite{Chapman:2016hwi} that the null joint contributions at the UV cutoff surfaces have no time dependence, so we need only to consider the joint at $r_{m}$. Assume that $k_1$ and $k_2$, the null vectors associated with two past null boundaries intersecting at $r_{m}$, are defined by
 \be\label{kk} k_1=\xi\left(-dt+\frac{dr}{f(r)}\right), \qquad k_2=\xi\left(dt+\frac{dr}{f(r)}\right),\ee
where $\xi$ is a normalization constant for null vectors. Following \cite{Lehner:2016vdi}, the joint term $a\!=\!\ln|-\frac{ k_1\cdot k_2}{2}|$ in the action (\ref{joint}) can be evaluated as
 \be\label{joint3} I_{jnt}=\frac{V_2}{8\pi G_4} \Bigg[- \log\bigl(\frac{f(r)}{\xi^2}\bigr) U(r)\Bigg]_{r=r_m}.\ee
Due to the implicit time dependence of $r_m$ with equation $\label{rmeq} \frac{t}{2}-{r}_{\infty}^{*}+r_{}^{*} (r_m)=0$, or
\be\label{rmt}
\frac{dr_m}{dt}=-\frac{f(r_m)}{2},
\ee
and ${dI_{jnt}}/{dt}=-\frac{f(r_{m})}{2} \,{dI_{jnt}}/{dr_m}$, the time evolution of this contribution is
\be
\frac{dI_{jnt}}{dt}=\frac{V_2}{16\pi G_4} \Bigg[U(r) \prt_{r} { f(r)} + f(r)\prt_{r} { U(r)} \log\bigl(\frac{f(r)}{\xi^2}\bigr) \Bigg] _{r=r_m},
\ee
after substituting from (\ref{fp}) we obtain
\bea \label{joint4}
\frac{dI_{jnt}}{dt}&\!\!\!\!=\!\!\!\!&\frac{V_2}{32\pi G_4}\Bigg[\frac{Q^3(Q\!+\!7r)\!+\!Q{r}^2(15Q\!+\!13{r})\!+\!4{r}^4\!-\!(Q\!-\!2{r}) (Q\!+\!r_h)^3}{(Q \!+\! {r})}\!-\! \frac{\left(2 r r_h \!+\!Q (3 r \!-\! r_h)\right) \beta^2}{2(Q \!+\! r)}  \nn \\
&\!\!\!\!+\!\!\!\!&\log\bigl(\frac{f({r})}{\xi^2}\bigr) (Q + 4 {r}) ({r} -  r_h) \left(\frac{3 Q^2 + {r}^2 + {r} r_h + r_h^2 + 3 Q ({r} + r_h)}{ (Q + {r})}-\frac{\beta^2}{2 (Q + {r})}\right) \Bigg] _{r=r_m}.
\eea
In the late time limit where ${r_m}$ reaches the horizon at $r_h$, we have
\be \label{joint5}
 \frac{dI_{jnt}}{dt}{\Bigg|}_{t >>t_c}= \frac{V_2}{8\pi G_4} \Bigg[\frac{3}{2} r_h (Q + r_h)^2 -  \frac{1}{4} r_h  \beta^2\Bigg]= T S,
 \ee
 where as expected in the extremal limit becomes zero. The counterterm action (\ref{cterm}) for parametrization of the past and future null boundaries is
 \be\label{cterm3} I_{ct}=2\,(I_{ct}^{future}+I_{ct}^{past})=\frac{V_2}{4\pi G_4}\left (\int_{{\epsilon}}^{r_{max}}+\int_{r_m}^{r_{max}}\right) \prt_{r} { U(r)} \log\left(\frac{\ell_{c}\xi\prt_{r} U(r)}{U(r)}\right) dr,
 \ee
here, we have used the affine parameter $\lambda=r/\xi$, where $\xi$ is a constant, so the total action with the counterterm does not depend on the parametrization of the null surfaces \cite{Carmi:2017jqz} and the expansion (\ref{amb}) takes the form
\be\label{amb4}
 \Theta=\frac{\xi\prt_{r} U(r)}{U(r)}.
\ee
Then from (\ref{rmt}), the time evolution of the counterterm action becomes
\be\label{cterm4}
\frac{dI_{ct}}{dt}=\frac{V_2}{8\pi G_4}\Bigg[ \frac{ (r- r_h) (Q + 4 r)\bigl(6 Q^2 + 2 r_h^2 - \beta^2 + 2 r_h r + 2 r^2 + 6 Q (r_h + r)\bigr)}{8 (Q + r)}\log{\frac{l_c \xi (Q + 4 r)}{2 r (Q + r)}}\Bigg]_{r={r_m}}.
\ee

Although the gowth rate of the joint and counterterm actions depend on the parametrization constant $\xi$ for null surfaces, but it is found that the rate of complexity from the total action (\ref{tact}) is independent of it. In other words, the contribution of the joint action for $\xi$ is eliminated by the one for counterterm action.
\subsubsection{Late time behavior}
\begin{figure}[H]
\centering
\includegraphics[width=10cm,height=7cm]{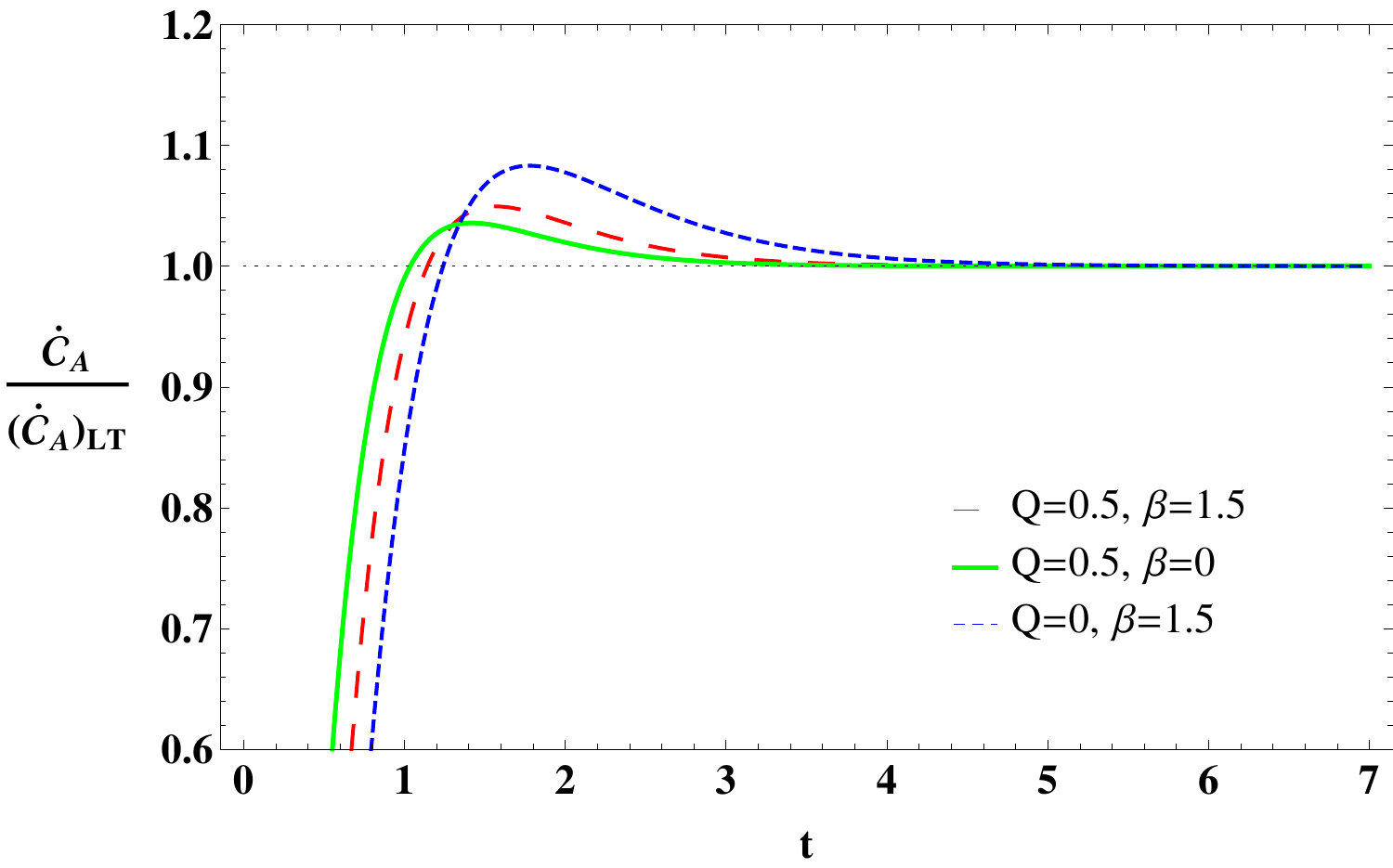}
\caption{The growth rate of complexity relative to its late time limit for different values of charge and momentum relaxation.}
\label{f2}
\end{figure}

To understand how $\dot{\cal{C}}_A$ violates the Lloyd's bound, we first investigate its late time behavior. Using the quantities in (\ref{mass4}) and (\ref{cpcd}), we obtain
\be \label{lt4}
\dot{\cal{C}}_A\Big|_{LT}\!=\!\frac{d I_{WDW}}{dt }{\Bigg|}_{t>> t_c}\!=\! \frac{V_2}{8\pi G_4}
\left[\frac{9}{2} Q r_h^2 + 2 r_h^3 + r_h (3 Q^2 -  \beta^2)\right]=2M- \mu q- \frac{V_2}{16\pi G_4} \left[Q^3-\frac12 Q \beta^2\right].
\ee
It is obvious that in the extremal limit ($r_h=0$), the late time behavior vanishes. Also, the result shows that not only the contribution of dilaton theory changes the late time limit of charged black holes in (\ref{cjlt}), see e.g. Ref.~\cite{Cai:2017sjv}, but also the presence of momentum dissipation makes a difference. The ratios $\dot{\cal{C}}_A/\big(\dot{\cal{C}}_A\big)_{LT}$ have plotted for different values of $Q$ and $\beta$ in Fig.~(\ref{f2}) where ``$LT$'' stands for the late time behavior given in equation (\ref{lt4}). As is obvious, $\dot{\cal{C}}_A$ violates the Lloyd's bound for this kind of black hole from above.
\subsection{ $AdS_5$ black holes}\label{sec32}

The extension of GR model \cite{Gubser:2009qt} in five dimensions to EMAD theories with momentum relaxation has been proposed in Ref.~\cite{Gouteraux:2014hca}. The EMD part comes from 10D type IIB string theory with geometry $AdS_5\times S^5$ as the near horizon limit of D3-branes \cite{Cvetic:1999xp}. With a similar trend, the translational invariance is broken by adding a kinetic term of axionic scalar fields $\psi_I$'s. The holographic charge transport and linear resistivity for charged $AdS_5$ black holes in this model have been reviewed in Refs.~\cite{Gouteraux:2014hca,Jeong:2018tua}. The bulk action of 5D EMAD theory is
\be \label{act5}
  I_{bulk} \!=\! {1 \over 16\pi G_5}\!\int\! d^5x \sqrt{-g} \left[ R\! -\! {1 \over 4} e^{4\phi} F_{\mu\nu}^2 \!-\!
    12 (\partial_\mu\phi)^2 \!+\! {1 \over L^2} (8e^{2\phi} \!+\!
      4e^{-4\phi}) \!-\!\frac{1}{2} \sum_{I=1}^3 {(\partial_\mu\psi_I)}^2 \right]\!=\!{1 \over 16\pi G_5}\!\int \!d^5 x \sqrt{-g} \mathcal{L}_2, \,
\ee
here the field content of this model is analogous to 4D case discussed in section \ref{sec31}. Analytic charged solution in this theory is given by the general ansatz (\ref{metric}) but with three-dimensional spatial space $d\vec{x}^2$, such that
\bea\label{p2}
&&  A(r) = \log {r \over L} + {1 \over 3} \log\left( 1 + {Q \over r^2} \right), \quad
  B(r) = -\log {r \over L} - {2 \over 3} \log\left( 1 + {Q \over r^2} \right), \quad \psi_I =\beta \delta_{I i} x^i,  \nn\\
&&  h(r) =1-  \frac{ \beta^2}{4 (Q + r^2)} -  \frac{(Q + r_h^2)^2 }{(Q + r^2)^2} \bigl(1 -  \frac{ \beta^2}{4 (Q + r_h^2)}\bigr),\quad \phi(r) = {1 \over 6} \log\left( 1 + {Q\over r^2} \right),\\
&&A_t(r) =\sqrt{2 Q \bigl(1 -  \frac{ \beta^2}{4 (Q + r_h^2)}\bigr)} (1-  \frac{ (Q + r_h^2)}{(Q + r^2)}).\nn
\eea
Again we set  $L=1$ without lose of generality. This is an asymptotically $AdS$ black hole which can be rewritten with a warped factor $w(r)$ as
\be
\label{ads50}
  ds^2 =e^{w(r)}  \left(-f(r)\, dt^2 +{U(r)} d\vec{x}^2 + \frac{1}{f(r)} \,{d}r^2 \right),
\ee
where
\be
w(r)=- {1 \over 3} \log\left( 1 + {Q \over r^2} \right),\quad f(r)=\frac{(r^2-r_h^2) (2 Q+ r_h^2 + r^2)}{Q + r^2}-\frac{  (r^2-r_h^2 )\beta^2}{4 (Q + r^2)},\quad {U(r)}=(Q + r^2).\nn
\ee

The mass of the black hole is given by
\be\label{mass5}
 M={ V_3 \over 16\pi G_5}\, 3\omega,\qquad \omega= (Q + r_h^2)^2\left(1-\frac{\beta^2}{4(Q + r_h^2)}\right),
\ee
where $V_3$ is the volume of the 3-dimensional flat space. In the limit $\beta=0$, it gives the mass parameter in Ref.~\cite{Gubser:2009qt} while in the limit $Q=0$, yields the mass of neutral black branes we obtained in \cite{Yekta:2020wup} for $d=4$. The chemical potential and charge of the black hole are computed from the gauge potential in (\ref{p2}) as
\be
\label{cpcd5}
\mu= \sqrt{2 Q \bigl(1 -  \frac{ \beta^2}{4 (Q + r_h^2)}\bigr)},\qquad
q ={ V_3 \over 8\pi G_5}  (Q + r_h^2) \sqrt{2 Q \bigl(1 -  \frac{ \beta^2}{4 (Q + r_h^2)}\bigr)}.
\ee
On the other hand, the temperature and entropy of the black hole are
\be \label{ST5}
T=\frac{1}{8 \pi} r_h (8 -  \frac{ \beta^2}{Q + r_h^2}),\qquad S=\frac{V_3}{4 G_5} r_h (Q + r_h^2)
\ee
so that one has
\be  \label{ST50}
T S=\frac{V_3}{8 \pi G_5} \frac{ r_h^2}{4} \bigl(8 (Q + r_h^2) - \beta^2\bigr).
\ee
\subsubsection{The growth rate of complexity}
Following the prescription in section \ref{sec2}, we obtain the growth rate of holographic complexity for charged $AdS_5$ black holes of 5D EMAD theory in the CA picture. The WDW patch in this case is also described by Fig.~(\ref{f1}) and we need to compute the total action (\ref{tact}) on it. In this regard, the contribution of the bulk action on three different regions $\RN{1}$, $\RN{2}$, and $\RN{3}$ for times $t>t_c$ is given by

\bea \label{bulk2} I_{bulk}&\!\!\!\!\!=\!\!\!\!\!&2\,\, \left(I_{bulk}^{\RN{1}}+I_{bulk}^{\RN{2}}+I_{bulk}^{\RN{3}}\right)\nn\\
&\!\!\!\!\!=\!\!\!\!\!&\frac{V_3}{8\pi G_5}\left[\int_{\epsilon}^{r_h}\left(\frac{t}{2}+{r}_{\infty}^{*}-r_{}^{*} (r)\right)+2\int_{r_h}^{r_{max}}\left({r}_{\infty}^{*}-r_{}^{*} (r)\right)+\int_{r_m}^{r_h}\left(-\frac{t}{2}+{r}_{\infty}^{*}-r_{}^{*} (r)\right) \right]\, \sqrt{-g}\, \mathcal{L}_2\, dr,\nn\\
&\!\!\!\!\!=\!\!\!\!\!&I_{bulk}^{0}+\frac{V_3}{8\pi G_5} \int_{\epsilon}^{r_m}\left(\frac{t}{2}+{r}_{\infty}^{*}-r_{}^{*} (r)\right) \, \sqrt{-g}\, \mathcal{L}_2\,  dr,
\eea
where $I_{bulk}^{0}$ is the independency of the time and $\mathcal{L}_2$ is the Lagrangian density of (\ref{act5}) which on the background (\ref{p2}) we have
\be
\sqrt{-g}{\mathcal{L}}_{2}= -  \frac{8 r \bigl(Q^3 + Q^2 (7 r^2-2 r_h^2 ) + Q (8 r^4-r_h^4 )+ 3 r^6 \bigr)}{3  (Q + r^2)^{2}}- \frac{2 r Q (Q + r_h^2)  \beta^2}{3(Q + r^2)^{2}}.
\ee
Thus the time derivative of (\ref{bulk2}) yields
\be
\frac{d I_{bulk}}{dt }=  \frac{V_3}{8\pi G_5} \left[- \tfrac{4}{3} Q r^2 -  r^4 -  \frac{2 Q (Q + r_h^2)^2}{3 (Q + r^2)} + \frac{Q (Q + r_h^2)  \beta^2}{6 (Q + r^2)}\right]_{\epsilon}^{r_{m}}.
\ee
Note that the time evolution of this result comes from the differential equation of $r_m$ for the past null junction given in (\ref{rmt}).

In order to find the extrinsic curvature for geometry (\ref{ads50}), we define the normal vector $n_{\mu}=(0,n_r,0,0,0)$ where $n_r=\sqrt{\frac{f(r)}{W(r)}}$ and for simplicity we choose $W(r)=e^{w(r)}$. Thus, from the definition in section \ref{sec2} one can achieve the following expression
 \be
K= \frac{4 f(r) \,U(r)\, W(r)' + W(r) \bigl(3 f(r)\, U(r)' + f(r)' \,U(r)\bigr)}{2  f(r)^{1/2}\,U(r)\,W(r)^{3/2}},
\ee
where prime is the derivative with respect to $r$. For times $t>t_c$ the boundary action (\ref{surf}) becomes
\be
\label{surf5} I_{surf}=I_{surf}^0+\frac{V_3}{4\pi G_5}\sqrt{h} K\,\left(\frac{t}{2}+{r}_{\infty}^{*}-r_{}^{*} (r)\right)\Big|_{r=\epsilon}.
 \ee
Therefore its time derivative gives
\be
 \frac{d I_{surf}}{dt}= \frac{V_3}{8\pi G_5 } \sqrt{h} { K  {\Bigg|}_{r=\epsilon}},
 \ee
in which
\bea
\sqrt{h}K&\!\!\!=\!\!\!&- \frac{2 \bigl(Q^2 (7 r^2-4 r_h^2 ) + 2 Q (7 r^4-r_h^4 - 3 r_h^2 r^2) + 3 r^2 (2 r^4-r_h^4)\bigr)}{3 (Q + r^2)}\nn\\
&\!\!\!-\!\!\!&\frac{ \bigl(6 r_h^2 r^2 - 9 r^4 + Q (4 r_h^2 - 7 r^2)\bigr) \beta^2}{12 (Q + r^2)}.
\eea

For the null joint contribution of intersecting the two past null boundaries at $r = r_m$ with outward-directed null vectors defined in (\ref{kk}) we have
\be
\label{joint5}
I_{jnt}=-\frac{V_3}{8\pi G_5 } \,\left[U(r)^{3/2} \,W(r)^{3/2} \log\Bigl(\frac{\bigl|W(r) f(r)\bigr|}{\xi^2}\Bigr)\right]_{r=r_m},
\ee
where the implicit time dependence of this term is through the Eq.~(\ref{rmt}). For the background functions in (\ref{ads50}), the joint term can then be evaluated as
\bea
\label{joint50}
\frac{dI_{jnt}}{dt}&=&\frac{V_3}{ 8\pi G_5 } \Bigg[\frac{ (Q + 3 r^2) (r^2 -  r_h ^2) (2 Q + r^2 + r_h^2)}{3 r^{1/3} (Q + r^2)^{4/3}} \, \log\Bigl(\frac{\bigl|W(r) f(r)\bigr|}{\xi^2}\Bigr) \nn \\
&+&\frac{8 Q^2 r^2 + 7 Q r^4 + 3 r^6 - (Q - 3 r^2)(2 Q+r_h^2)  r_h^2}{3 r^{1/3} (Q + r^2)^{4/3}}\nn \\
&-&\left(\frac{(Q + 3 r^2) (r^2 -  r_h ^2)}{12 r^{1/3} (Q + r^2)^{4/3}} \, \log\Bigl(\frac{\bigl|W(r) f(r)\bigr|}{\xi^2}\Bigr) + \frac{Q(r^2 -  r_h ^2)+3 r^2(Q +r_h^2)}{12 r^{1/3} (Q + r^2)^{4/3}}\right)  \beta^2
\Bigg]_{r=r_m},
\eea
in which
\be
 \log\Bigl(\frac{\bigl|W(r) f(r)\bigr|}{\xi^2}\Bigr) =\log\bigl(\frac{r^{2/3} (r^2- r_h^2 ) (2 Q + r^2 + r_h^2 - {\beta^2} /4)}{\xi^2 (Q + r^2)^{4/3}}\bigr).
\ee
In the late time limit one can find that the growth rate of the joint action is proportional to the product of the temperature and entropy
\be
 \frac{dI_{jnt}}{dt}{\Bigg|}_{t >>t_c}= \frac{V_3}{ 8\pi G_5 } \Bigg[2 r_h^2 (Q^2 + r_h^2) -  \tfrac{1}{4} r_h^2  \beta^2\Bigg]= T\,S,
 \ee
where in the extremal limit vanishes.

The boundary counterterm (\ref{cterm}) requires evaluating the expansion scalar (\ref{amb}) in the null boundaries of the WDW patch, thus with the parametrization $\lambda=r/\xi$ we obtain
\be
\label{amb5}
\Theta=\frac{3 \xi}{2} \left(\frac{U(r)'}{U(r)} + \frac{W(r)'}{W(r)}\right).
\ee
Similar to the action (\ref{cterm3}) the evaluation for $\epsilon$ vanishes and the UV regulator surface at $r_{max}$ only contributes a fixed constant, therefore the time dependence comes only from the term evaluated at the meeting point $r_m$. Now, by virtue of dynamical equation (\ref{rmt}) we can determine time evolution of the counterterm action as
\be \label{cterm5}
\frac{dI_{ct}}{dt}=\frac{V_3}{16\pi G_5} \Bigg[\frac{( r^2- r_h^2 ) (Q + 3 r^2) (2 Q + r_h^2 + r^2-\beta^2 /4 )}{(Q + r^2)}\,\log\left( \frac{l_c \xi (Q + 3r^2)}{ r(Q + r^2)}\right) \Bigg]_{r=r_m}.
\ee

\subsubsection{Late time behavior}
\begin{figure}[H]
\centering
\includegraphics[width=10cm,height=7cm]{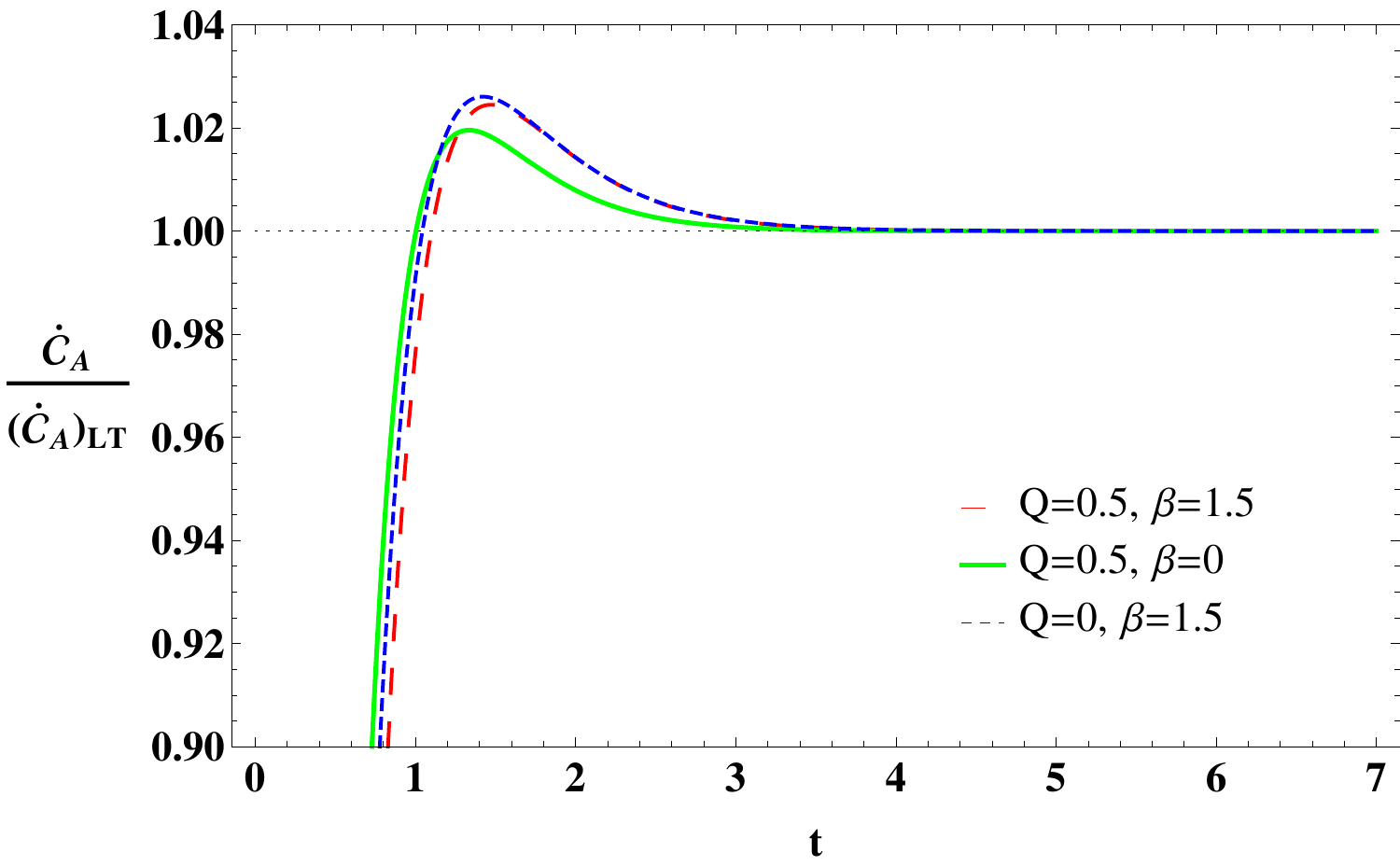}
\caption{The growth rate of complexity relative to its late time limit for different values of the charge and momentum relaxation.}
\label{f3}
\end{figure}

The rate of the complexity at very late times for the total action (\ref{tact}) is given by
\be
\label{lt5}
\frac{d I_{WDW}}{dt }{\Bigg|}_{t>>t_c}= \frac{V_3}{8\pi G_5 }
\Bigg[4 Q r_h^2 + 3 r_h^4 -  \tfrac{3}{4} r_h^2  \beta^2\Bigg],
\ee
where by using the quantities in Eqs.~(\ref{mass5}) and (\ref{cpcd5}) we can rewrite it as
\be
\label{lt50}
\dot{\cal{C}}_A\Big|_{LT}=\frac{d I_{WDW}}{dt }{\Bigg|}_{t>>t_c}=2M - \mu q- \frac{V_3}{16\pi G_5 }\left[2Q^2-\frac12 Q \beta^2\right].
\ee
As is obvious, the momentum relaxation and dilatonic field terms have explicit contribution in the Lloyd's bound due to the charge parameter $Q$. The ratios $\dot{\cal{C}}_A/\big(\dot{\cal{C}}_A\big)_{LT}$ have plotted for different values of $Q$ and $\beta$ in Fig.~(\ref{f3}). As seen in figure, $\dot{\cal{C}}_A$ violates the Lloyd's bound for this kind of black hole from above.

\subsection{Warped $AdS_{d+1}$ black holes}
Since the 4D and 5D EMAD theories used to studied had analytic solutions, it would be of interest to investigate a generalized theory in arbitrary dimensions. This model has been proposed in Ref.~\cite{Gouteraux:2014hca} which still allows analytic charged solutions. The holographic characteristics of strange metals for this model have been studied in Refs.~\cite{Gouteraux:2014hca,Jeong:2018tua}.

Let us consider a general $(d+1)$-dimensional EMAD theory given by the action (\ref{pact}) with the following Lagrangian density
\be \label{bulkd}
\mathcal{L}_3 =  R -  \frac{1}{2}(\partial_\mu\phi)^2 + V(\phi) -  \frac{1}{4} Z(\phi) F_{\mu\nu}^2  -\frac{1}{2} \sum_{i=1}^{d-1}{(\partial_\mu\psi_I)}^2,
\ee
where the functions $Z(\phi)$ and $V(\phi)$ are assumed to have the following specific forms \cite{Gouteraux:2014hca}
\be
\label{ZV}
Z(\phi)=e^{-(d-2) \delta  \phi},\qquad V(\phi) = V_1 e^{\frac{ (d-2 ) (d-1)  \delta^2-2 }{2 (d-1)  \delta}\phi }  + V_2 e^{-\frac{2 }{(d-1)  \delta}\phi }  + V_3  e^{(d-2)  \delta \phi }.
\ee
Here $V_1$, $V_2$, and $V_3$ are three constant parameters given by
\bea
 V_1 &\!\!=\!\!&\frac{8 (d-2 ) (d-1)^3  \delta^2}{\bigl( (d-1) (d-2)\delta^2+2\bigr)^2},\qquad V_2=\frac{(d-2)^2 (d-1)^2 \bigl( d(d-1)  \delta^2-2\bigr)  \delta^2}{\bigl( (d-1)(d-2)   \delta^2+2\bigr)^2},\nn\\
 V_3&\!\!=\!\!&\frac{4d (d-1) - 2 (d-1)^2(d-2)^2  \delta^2}{\bigl(  (d-1)(d-2) \delta^2+2\bigr)^2},
\eea
where $\delta$ is a free parameter. It should be noticed that for $d=3$ by choosing $\delta=\sqrt{\frac13}$ and redefinition $\phi\rightarrow -\sqrt{3}\phi$ the bulk action (\ref{bulkd}) reduced to 4D EMAD theory given in (\ref{act4}), while for $d=4$, $\delta=\sqrt{\frac16}$ and redefinition $\phi\rightarrow -\sqrt{24}\phi$, it yields the 5D theory described by (\ref{act5}).

The Eqs. (\ref{eqs}) admit an analytical charged black hole solution defined by the ansatz (\ref{metric}) with the following quantities
\bea \label{pd}
&&  A =\frac{1}{2} \log\Bigl(r^{2 - \frac{4}{ (d-1)(d-2) \delta^2+2}} (Q + r^{d-2})^{\frac{4}{(d-2) \left((d-1)(d-2)   \delta^2+2\right)}}\Bigr),\quad \psi_I =\beta \delta_{I i} x^i,\nonumber\\
&& B=\frac{1}{2} \log\bigl(r^{\frac{4 (d-2)}{ (d-1)(d-2) \delta^2+2}-2} (Q + r^{d-2})^{-\frac{4}{ (d-1)(d-2) \delta^2+2}}\bigr),\quad \phi(r)=- \frac{2 ( d-1)  \delta}{ (d-1)(d-2) \delta^2+2} \log\left(1 + \frac{Q }{r^{d-2}}\right), \nonumber\\
&&  h(r) = \bigl(1 -  \frac{\beta^2}{2 ( d-2) (Q + r^{ d-2})^{\frac{2}{ d-2}}}\bigr) -  \frac{ (Q + r_h^{ d-2})^{\frac{d}{ d-2}} }{ (Q + r^{ d-2})^{\frac{d}{ d-2}}} \bigl(1 -  \frac{\beta^2}{2 ( d-2) (Q + r_h^{ d-2})^{\frac{2}{ d-2}}}\bigr),\\
&& A_t(r) =\sqrt{\frac{d Q}{d-2}} (Q + r_h^{ d-2})^{-\frac{ d-4}{2 ( d-2)}} \sqrt{1 -  \frac{ \beta^2}{2 ( d-2) (Q + r_h^{ d-2})^{\frac{2}{ d-2}}}} (1 -  \frac{Q + r_h^{d-2}}{Q + r^{ d-2}}).\nonumber
\eea
Here, $r_h$ is the location of the event horizon and we have set here $L=1$. In analogy to the previous sections, we can recast this geometry as an asymptotically $AdS_{d+1}$ black hole
\be
\label{adsd}
   ds^2 =e^{w(r)}  \left(-f(r)\, dt^2 +{U(r)}\, d \Sigma_{d-1}^2 + \frac{1}{f(r)} \,\, \mathrm{d}r^2\right),
\ee
where $W(r)=e^{w(r)}$ and the corresponding functions are obtained as
\bea
&&f(r)=g(r)^{-\frac{2 (d-1)}{( d-2) \bigl( (d-1)(d-2) \delta^2+2\bigr)}} \mathcal{F}(r),\quad U(r)=g(r)^{\frac{d-1}{2( d-2) \bigl( (d-1)(d-2) \delta^2+2\bigr)}},\nonumber\\
&&W(r)=g(r)^{-\frac{2 (d-3)}{( d-2) \bigl( (d-1)(d-2) \delta^2+2\bigr)}},\quad g(r)=1 + \frac{Q }{r^{d-2}},  \\
&&\mathcal{F}(r)= r^2 \Bigl(- \frac{r_h^d}{r^d} g(r_h)^{\frac{4 ( d-1)}{( d-2) \bigl( (d-1)(d-2) \delta^2+2\bigr)}}+ g(r)^{\frac{4 ( d-1)}{( d-2) \bigl( (d-1)(d-2) \delta^2+2\bigr)}}\Bigr)- \frac{\beta^2}{2 (d-2)} (1 -\frac{ r_h^{d-2}}{r^{d-2}}) .\nn
\eea

In this paper we will consider the case with $\delta=\sqrt{\frac{2}{d(d-1)}}$ which corresponds to the GR model in higher dimensions and the IR geometry is conformal to $AdS_2 \times R^{d-1}$ \cite{Gouteraux:2014hca}. It is noticed that $\delta=0$ corresponds to a Reissner-Nordstrom $AdS$ black hole but in general one can consider an arbitrary value of $\delta\in\left[0,\sqrt{\frac{2}{d(d-1)}}\right]$. Further discussion for $\delta=\frac{1}{\sqrt{3}}$ could be found in Ref.~\cite{Jeong:2018tua}. Due to this fact, the mass of the black hole is given by
\be \label{massd}
M={V_{d-1} \over 16\pi G_{d+1}}  (d-1) \, \omega,\quad \omega=(Q + r_h^{d-2})^{ \frac{d}{d-2}} \bigl(1-\frac{ \beta^2}{2(d-2) (Q + r_h^{d-2})^{ \frac{2}{d-2}}}\bigr),
\ee
where $V_{d-1}$ is the volume of ($d-1$)-dimensional flat space with the line element $d \Sigma_{d-1}^2$. The chemical potential and charge of the black hole can be found from $A_{t}(r)$ in (\ref{pd}) as
\bea \label{cpcdd}
q &\!\!\!=\!\!\!& {V_{d-1} \over 16\pi G_{d+1}} \sqrt{d (d-2) Q  (Q + r_h^{ d-2})^{\frac{ d}{( d-2)}} \left(1 -  \frac{ \beta^2}{2 ( d-2) (Q + r_h^{ d-2})^{\frac{2}{ d-2}}}\right)}\,,\nonumber\\
\mu &\!\!\!=\!\!\!& \sqrt{\frac{d Q}{d-2}\,(Q + r_h^{ d-2})^{-\frac{ d-4}{( d-2)}}\left(1 -  \frac{ \beta^2}{2 ( d-2) (Q + r_h^{ d-2})^{\frac{2}{ d-2}}}\right)}\,.
\eea
The temprature and entropy of the black hole at the event horizon are given by
\be \label{STd}
T =\frac{r_h^{ d/2 -1}}{4 \pi }\biggl( d (Q + r_h^{ d-2})^{-\frac{d-4}{2( d-2)}} -  \frac{1}{2} (Q + r_h^{ d-2})^{-\frac{d}{2 (d-2)}} \beta^2 \biggr),\quad S =\frac{V_{d-1}}{4 G_{d+1} } r_h^{ d/2 -1} (Q + r_h^{ d-2})^{\frac{d}{2 ( d-2)}},
\ee
such that
\be \label{STpd}
T S =\frac{V_{d-1}}{16\pi G_{d+1} } \bigg[ d r_h^{ d-2} (Q + r_h^{d-2})^{\frac{2}{ d-2}} -  \tfrac{1}{2} r_h^{ d-2}  \beta^2\bigg].
\ee
 \subsubsection{The growth rate of complexity}
The time evolution of holographic complexity for $AdS$ black holes of generalized EMAD theory on the WDW patch shown in Fig.~(\ref{f1}) is determined by the rate of action (\ref{tact}) via CA proposal. Therefore, for the bulk action (\ref{pact}) on the right sector of WDW patch denoted by regions $\RN{1}$, $\RN{2}$ and $\RN{3}$ we have
\bea \label{bulk3} I_{bulk}(t>t_c)=2 \left(I_{bulk}^{\RN{1}}+I_{bulk}^{\RN{2}}+I_{bulk}^{\RN{3}}\right)
=I_{bulk}^{0}+\frac{V_{d-1}}{8\pi G_{d+1}} \int_{\epsilon}^{r_m}\left(\frac{t}{2}+{r}_{\infty}^{*}-r_{}^{*} (r)\right) \, \sqrt{-g}\, \mathcal{L}_3\, dr,
\eea
where similar to the previous subsections, $I_{bulk}^{0}$ is independent of the time and $\mathcal{L}_3$ is the Lagrangian density in Eq.~(\ref{bulkd}) evaluated on the background (\ref{pd}). Thus, the growth rate of the bulk contribution becomes
\be \label{bulk4}
\frac{d I_{bulk}}{dt }=  \frac{V_{d-1}}{8\pi G_{d+1} }\Bigg[-\frac{\bigl(\frac{( d-2) Q}{2 ( d-1)} + r^{ d-2}\bigr) }{(Q + r^{ d-2})^{-\frac{2}{ d-2}}} + \frac{d Q (Q + r_h^{ d-2})^{\frac{d}{ d-2}}}{2(d-1) (Q + r^{ d-2})} + \frac{d Q (Q + r_h^{ d-2})  \beta^2}{4 ( d-1) (d-2) (Q + r^{ d-2})}\Bigg]_{\epsilon}^{r_{m}}.
\ee

The extrinsic curvature for the boundary GHY surface term calculated for the normal vector with $n_r=\sqrt{\frac{f(r)}{W(r)}}$ is obtained from (\ref{adsd}) as
\be \label{surfd}
\sqrt{h}K=\frac{1}{2}\,U(r)^{\frac{( d-1)}{2} }W(r)^{\frac{( d-1)}{2}} f(r)' + \frac{1}{2}\,f(r)\, U(r)^{\frac{( d-1)}{2}}\,W(r)^{\frac{(d-3)}{2}} \left(( d-1) W(r) U(r)' + d\, U(r)  W(r)'\right),
\ee
which from (\ref{pd}) it yields
\bea \label{surfd1}
\sqrt{h}K&\!\!=\!\!&-\,d (Q + r^{ d-2})^{\frac{d}{ d-2}} +  \frac{d^2 Q}{2 ( d-1) (Q + r^{ d-2})^{-\frac{2}{ d-2}}} +  \frac{d \bigl(\frac{( d-2) Q}{ d-1} + r^{ d-2}\bigr) (Q + r_h^{ d-2})^{\frac{d}{ d-2}}}{2 (Q + r^{ d-2})}\nonumber
 \\
&&+\frac{Q \bigl((2d-1) r^{ d-2} - d r_h^{ d-2}\bigr)  \beta^2}{4 ( d-1) (Q + r^{ d-2})} + \frac{r^{ d-2} \bigl(2 ( d-1) r^{ d-2} - d r_h^{ d-2}\bigr) \beta^2}{4 ( d-2)  (Q + r^{ d-2})}.
\eea
As noticed in Eq.~(\ref{surf3}) the boundary contribution coming from the time-like surface at the UV cutoff regulator $r=r_{max}$ yields a fixed constant, i.e., they do not contribute to the time derivative of the action, and with affinely-parametrized null normals
($\kappa=0$), the null surface term vanishes. Thus, we only need to consider the boundary surface at the future singularity which is given by
\be \label{surfd2}
 \frac{d I_{surf}}{dt}= \frac{V_{d-1}}{ 8\pi G_{d+1}}   \sqrt{h}  K  {\Bigg|}_{r=\epsilon}.
 \ee

The final result for the joint action at the meeting point $r = r_m$, shown in Fig.~(\ref{f1}), is given by
\be \label{jointd}
I_{jnt}= -\frac{V_{d-1}}{ 8\pi G_{d+1}} \left[ U(r)^{\frac{d-1}{2}} W(r)^{\frac{d-1}{2}}\log\bigl(\frac{f(r) W(r)}{\xi^2}\bigr)  \right]_{r=r_m},
\ee
where $\xi$ is the normalization constant appearing in the null normals (\ref{kk}), i.e., $k\cdot \prt_t \big|_{r\rightarrow \infty}=\pm\xi$. According to the time dependence of $r_m$ through Eq.~(\ref{rmt}) the evolution of this action has the following form
\bea \label{tjd}
\frac{dI_{jnt}}{dt}&\!\!\!=\!\!\!&\frac{V_{d-1}}{ 8\pi G_{d+1} } \Bigg[\frac{ (Q+r^
{d-2}  )^{\frac{2}{d-2}} \bigl((d-2) Q + 2 (d-1) r^{d-2}\bigr) \bigl(2 + (d-1) \log(\frac{\mathcal{F}(r)}{\xi^2}) \bigr)}{4 (d-1) }
 \\
&\!\!\!-\!\!\!&\frac{(d-2) \bigl(Q - (d-1) r^{d-2}\bigr) (Q+r_h^{d-2})^{\frac{d}{d-2}}}{2 (d-1 ) (Q+r^{d-2})} +\frac{\bigl((1 + \frac{Q }{r^{d-2}}) -  d (1 + \frac{Q }{r_h^{d-2}}) \bigr)  (r\,r_h)^{d-2} \beta^2}{4 (d-1) (Q+r^{d-2})} \nonumber
 \\
&\!\!\!-\!\!\!& \frac{\bigl((d-2) Q + 2 (d-1) r^{d-2}\bigr) (Q+r_h^{d-2})^{\frac{d}{d-2}}\log(\frac{\mathcal{F}(r)}{\xi^2})}{4(Q+r^{d-2}) }\nonumber
 \\
&\!\!\!-\!\!\!& \frac{\bigl((d-2 ) Q + 2 (d-1) r^{d-2}\bigr) ( r^{d-2} - r_h^{d-2})  \beta^2 \log(\frac{\mathcal{F}(r)}{\xi^2})}{8 (d-2) (Q+r^{d-2})}\Bigg]_{r=r_m},\nonumber
\eea
in which
\be
\mathcal{F}(r) =r^{\frac{d-2}{d-1}} (Q+ r^{d-2})^{-\frac{d}{d-1}}\left( ( Q+ r^{d-2})^{\frac{d}{d-2}}-   (Q+ r_h^{d-2})^{\frac{d}{d-2}}  - \frac{ \beta^2}{2(d-2)}  \Bigl(r^{d-2}-r_h^{d-2}\Bigr)\right).
\ee
As shown before, in the late time limit $r_m$ coincides with the event horizon $r_h$, then from Eq.~(\ref{STpd}) the joint term (\ref{tjd}) leads to
\be
\frac{dI_{jnt}}{dt}{\Bigg|}_{t >>t_c}=\frac{V_{d-1}}{16\pi G_{d+1}} \bigg[ d r_h^{ d-2} (Q + r_h^{d-2})^{\frac{2}{ d-2}} -  \frac{1}{2} r_h^{ d-2}  \beta^2\bigg]=TS.
 \ee

By fixing the parametrization of null generators as $\lambda=r/\xi$, the boundary counterterm action for the null boundaries is defined by the following expansion parameter
\be \label{ambd}
\Theta=\frac{(d-1) \xi}{2} \left(\frac{U(r)'}{U(r)} + \frac{W(r)'}{W(r)}\right)=\frac{( d-2) Q + 2 (d-1) r^{d-2}}{2 r (Q + r^{ d-2})} \xi,
\ee
such that the time evolution of counterterm action becomes
\bea \label{ctermd}
\frac{dI_{ct}}{dt}&=&\frac{V_{d-1}}{16\pi G_{d+1}}\Bigg[\frac{( d-2) Q + 2 (d-1) r^{d-2}}{(Q + r^{ d-2})}
\biggl( \frac{ \bigl(  (Q + r^{ d-2})^{\frac{d}{ d-2}}- (Q + r_h^{ d-2})^{\frac{d}{ d-2}}\bigr)}{2}\nonumber \\
&&\qquad \qquad- \frac{( r^{ d-2}- r_h^{ d-2} )}{4 ( d-2)} \beta^2 \biggr)\times  \log\biggl( \frac{( d-2) Q + 2 (d-1) r^{d-2}}{2 r (Q + r^{ d-2})} l_c \xi\biggr)
\Bigg]_{r=r_m}.
\eea
\subsubsection{Late time behavior}
Now we are ready to calculate the rate of holographic complexity at large times for the total action $I_{WDW}$ by sum over the above results according to Eq.~(\ref{tact}) as
\bea \label{ltd1}
\dot{\cal{C}}_A\Big|_{LT}&=&\frac{d I_{WDW}}{dt }{\Bigg|}_{t>>t_c}\\
&=&\frac{V_{d-1}}{ 16\pi G_{d+1} }
\Bigg[-(d-2 ) Q^{\frac{d}{d-2}} + (Q + r_h^{ d-2})^{\frac{2}{ d-2}} \left(( d-2) Q + 2 ( d-1) r_h^{ d-2}\right) -  \frac{( d-1) r_h^{ d-2}  \beta^2}{( d-2)}\Bigg].\nn
\eea
Substituting from (\ref{massd}) and (\ref{cpcdd}) in Eq.~(\ref{ltd1}) we obtain
\be \label{ltd2}
\dot{\cal{C}}_A\Big|_{LT}=2 M -  \mu q - \frac{V_{d-1}}{ 16\pi G_{d+1} }
\Bigg[ ( d-2) Q^{\frac{d}{ d-2}} - \frac{1}{2} Q \beta^2\Bigg].
\ee

In the absence of momentum relaxation term this result reduces to the following Lloyd's bound for the $(d+1)$-dimensional charged $AdS$ black holes in the generalized GR model
\be \label{ltd3}
\dot{\cal{C}}_A\Big|_{LT}=2M_0-\mu_0 q_0-D,\qquad D=\frac{V_{d-1}}{ 16\pi G_{d+1} }( d-2) Q^{\frac{d}{d-2}},
\ee
where the quantities $M_0$, $\mu_0$ and $q_0$ are given by the relations (\ref{massd}) and (\ref{cpcdd}) when $\beta=0$. It should be also noted that both of Eqs.~(\ref{ltd2}) and (\ref{ltd3}) vanish in the extremal limit $r_h=0$.

\section{Conclusions and outlook}\label{sec4}
One of the main implications in the context of $AdS/CFT$ duality, which relates the physics of the black holes in gravity side to the information theory in dual field theory, is the notion of quantum complexity. In other words, holographic methods are powerful tools to study the properties of strongly correlated systems by mapping them to the dual weakly interacting systems. In this regard, the GR models are of great importance. In this paper, we investigated the holographic complexity for the charged $AdS$ black holes in these models when the translational invariance is broken in diverse dimensions. Following the prescription in \cite{Carmi:2017jqz} with the inclusion of joint and counterterm actions for null boundaries \cite{Lehner:2016vdi}, we studied the growth rate of complexity on the WDW patch shown in Fig.~(\ref{f1}) for a modification of GR model to holographic theory containing axion fields which yields the momentum relaxation (EMAD theories).

We have shown that though the time evolution of the joint and counterterm for null surfaces depend separately on the normalization condition $\xi$, which is needed to remove the reparametrization ambiguity of null boundaries, the sum of the two terms in $\dot{\cal{C}}_A=dI_{WDW}/dt$ is independent of $\xi$. We examined the standard calculations for the three most used models, i.e., 4, 5, and $(d+1)$-dimensional EMAD theories with single-horizon charged $AdS$ black holes. According to the curves in Figs.~(\ref{f2}) and (\ref{f3}), the growth rate of holographic complexity is finite for early times in each case although it violates the Lloyd's bound. However, at late times it saturates the corresponding bound which is not $2M$. In particular, the results for GR model and EMAD theories in arbitrary dimensions are given by Eqs. (\ref{ltd3}) and (\ref{ltd2}), respectively.

To have a comparison of our results with other literature, we have summarized the late time behavior of $\dot{\cal{C}}_A$ for neutral and charged $AdS$ black holes in different theories in Tab.~(\ref{tab1}). It is shown that for neutral black holes of mass $M$, the Lloyd's bound is equal $2M$ even by adding the momentum relaxation term or higher order curvature corrections. For charged solutions with single horizon, we have the contribution of chemical potential term too. When there are non-linear expressions of dilatonic or gauge fields, we have extra term in the bound due to these corrections  \footnote{In the table we have set $V_{d-1}=16\pi G$ and the constants are $C=b_0^{1/2} Q^{3/2} \frac{\Gamma\left(\frac14\right)\Gamma\left(\frac54\right)}{3\Gamma\left(\frac12\right)}$, $D=\frac{Q^2 e^{2\phi_0}}{2M}$ and $C_0=\frac{g^2 r_{\star}^3 (\beta \gamma+4\kappa)}{10 \pi}$.}. Finally, we have shown that in the presence of momentum relaxation term in EMAD theory we have a correction of $Q$ and $\beta$ which are given in the last two rows in Tab.~(\ref{tab1}).

\begin{table}[H] 
 \caption[]{Late time limit of $\dot{\cal{C}}_A$ for single horizon black holes in different theories.}
\centering
\begin{tabular}{|c|c|}\hline
 Model & $\big(\dot{\cal{C}}_A\big)_{LT}$  \T\B \\ \hline
 Neutral $AdS$ BHs of Einstein theory \cite{Brown:2015lvg,Carmi:2017jqz} &  $2M$ \T\\  \hline
  Neutral $AdS$ BHs of EMA theory \cite{Yekta:2020wup} & $2M$ \T\\ \hline
  Neutral $AdS$ BHs of higher curvature theories \cite{Jiang:2018pfk,Fan:2019aoj,Ghodsi:2020qqb} & $2M$ \T\\ \hline
 Anisotropic BHs of Mateos and Trancanelli theory \cite{HosseiniMansoori:2018gdu} & $2M$ \T\\ \hline
Extremal RN-$AdS$ BHs of EM theory \cite{Brown:2015bva}& $ 2M-\mu \, q $ \T\\ \hline
Charged $AdS$ BHs of Born-Infeld theory \cite{Cai:2017sjv}& $ 2M-\mu \, q-C $ \T\\ \hline
 Charged $AdS$ BHs of EMD theory \cite{Cai:2017sjv}\cite{Goto:2018iay} & $2M-\mu\,q-D   $\T \\ \hline
Charged $AdS$ BHs of Horndeski theory \cite{Feng:2018sqm} &  $ 2M-\mu\,q-C_0   $ \T\\ \hline
Charged $AdS$  BHs of generalized GR theory  & $  2 M -  \mu \, q -  ( d-2) Q^{\frac{d}{ d-2}}$ \T\\ \hline
Charged $AdS$ BHs of EMAD theory & $ 2 M -  \mu \, q - ( d-2) Q^{\frac{d}{ d-2}} + \frac{1}{2} Q \beta^2 $\T\B\\\hline
\end{tabular}
\label{tab1}
\end{table}

 It would be also of interest to study the holographic complexity for other generalization of the holographic axion models in \cite{Andrade:2013gsa,Davison:2015bea} by considering a special gauge-axion higher derivative term \cite{Gouteraux:2016wxj,Wang:2021jfu} of the form
\be S=\frac{1}{16\pi G_4}\int d^4 x \sqrt{-g} \left[R-2\Lambda-V(X)-\frac14\left(1+\mathcal{K}Tr[X]\right) F_{\mu\nu}F^{\mu\nu}\right],\ee
where $X^{I}=\beta \delta_{i}^{I}x^{i}$ are the axionic scalar fields, or other holographic axion models in \cite{Baggioli:2016oqk,Baggioli:2021xuv}.

\section*{Acknowledgment}
The authors would like to thank M. R. Mohammadi Mozaffar for his valuable comments and discussion.

   \end{document}